\documentclass[12pt,english]{iopart}
\usepackage{amssymb}
 \expandafter\let\csname equation*\endcsname\relax
  \expandafter\let\csname endequation*\endcsname\relax
\usepackage{amsmath}
\usepackage{xcolor}
\usepackage{graphicx}
\usepackage{braket}
\usepackage{mathtools}
\usepackage{iopams}

\begin{document}
\title[]{Stability and Hierarchy of Quasi-Stationary States: Financial Markets as an Example}
\date{March, 2015}
\author{Yuriy Stepanov$^1$, Philip Rinn$^2$, Thomas Guhr$^1$, Joachim Peinke$^2$ and Rudi Sch\"afer$^1$}
\address{$^1$ Faculty of Physics, University of Duisburg-Essen,  Duisburg, Germany}
\address{$^2$ Institute of Physics and ForWind, Carl-von-Ossietzky University Oldenburg, Germany}
\ead{yuriy.stepanov@uni-due.de}

\begin{abstract}
We combine geometric data analysis and stochastic modeling to describe the collective dynamics of complex systems. 
As an example we apply this approach to financial data and focus on the non-stationarity of the market correlation structure. We identify the dominating variable and extract its explicit stochastic model. This allows us to establish a connection between its time evolution and known historical events on the market. We discuss the dynamics, the stability and the hierarchy of the recently proposed quasi-stationary market states. 
\end{abstract}
\maketitle

\section{Introduction} \label{sec:intro}

\emph{Big data} is the buzzword of recent years, reflecting an ever increasing amount of electronically available data that demands analysis and interpretation.
Our focus is on complex dynamical systems such as financial markets, where huge data sets exist in the form of multivariate time series.
The dynamical behavior of such systems may reduce their complexity by self-organization \cite{Friedrich2011}. System variables, which are measured as single time series, couple together to a few dominating variables, which accurately describe the system dynamics and allow for predictions. The self-organization may produce patterns in observed data which are generally difficult to uncover. 
A wide range of data analysis techniques is available and widely used, including graph theoretical information filtering \cite{nicolo,assettrees,Mizuno2006336,rosario2,Tumminello201040,Tumminello26072005}, data clustering \cite{Pozzi:2013aa,rosario,ms,kmeans,Steinbach2000,dbht} and geometric approaches \cite{Hotelling,pearson1901lap,NLDR,eigen,eigen2}. 
All these techniques are based on a similarity measure between the data points. There is a major disadvantage in this approach: The time information of the measured data is neglected. Thus, the system dynamics is not explicitely taken into account. 
On the other hand, dynamical variables of complex systems have been successfully described by stochastic processes \cite{Friedrich2000-1,Friedrich2011,Renner2001}. In this description the variables evolve in time according to deterministic dynamics, which gives access to system stability and fixed points and is exposed to generally non-trivial stochastic fluctuations.
Here, we combine the data set analysis with stochastic methods in order to capture the full dynamics of the system. We apply our approach to stock market data. Similar techniques have proven successful in the description of complex dynamical systems  \cite{philip,PhysRevE.84.031103,PhysRevE.61.R4691}.
The paper is organized as follows: We present the data set and perform a geometric data analysis to uncover the dominating variable in Sec.~\ref{sec:datanal}. In Section \ref{sec:msperiod} we identify the quasi-stationary states of the financial market following Ref.~\cite{ms}. We draw connections to known historical events. We present the stochastic analysis in Sec.~\ref{sec:stochAnal} and discuss our results in Sec.~\ref{sec:results}.

\section{Analyzed Data} \label{sec:datanal}

In Sec. \ref{sec:quanti} we introduce our data set and the analyzed quantities. We perform a geometric analysis of the data in Sec. \ref{sec:pca}.
\subsection{Observed Quantities} \label{sec:quanti}

We analyze  daily adjusted closing stock prices $S_i(t)$ $i=1,...,K$ of the $K$ companies in the S$\&$P500 Index over the period of 21 years ranging from early 1992 to the end of 2012. The data is freely available at \texttt{finance.yahoo.com}. To measure the correlations, we use the daily returns
\begin{equation} \label{eq:return}
r_i(t)=\frac{S_i(t+1)-S_i(t)}{S_i(t)}
\end{equation} and normalize them locally \cite{locnorm}, to smooth out trends on very short times. We measure the time $t$ in trading days. We then calculate the $K\times K$ correlation matrices $C(t)$ by averaging over a time window of $T=42$ which is moved in one-day steps through the data. The elements of $C(t)$ are the Pearson correlation coefficients
\begin{equation}
C_{ij}(t)=\frac{\braket{r_{i}r_{j}}_{T}(t)-\braket{r_{i}}_{T}(t)\braket{r_{j}}_{T}(t)}{\sigma^{(T)}_{i}(t)\sigma^{(T)}_{j}(t)}.
\end{equation} Here $\sigma^{(T)}_{i}(t)=\sqrt{\braket{r_{i}^{2}}_{T}(t)-\braket{r_{i}}_{T}^{2}(t)} 
 $ is the time-dependent volatility and the sample average 
\begin{equation}
\braket{f}_T(t)=\frac{1}{T}\overset{t}{\underset{s=t-T+1}{\sum}}f(s)
\end{equation} of a quantity $f(t)$ is evaluated over the $T$ data points before $t$.  We note that in contrast to the stock prices $S_i$ and price returns $r_i$, the correlation coefficients $C_{ij}(t)$ are bounded quantities. All together we obtain $N=5169$ correlation matrices. 
 The correlation matrices calculated on the short intervals $T$ are noisy.  We reduce the noise by averaging over the correlation coefficients which yields the mean correlation coefficient 
\begin{equation}
\bar{c}(t)=\braket{C(t)}_{ij} \label{eq:meancor}.
\end{equation} Here $\braket{...}_{ij}$ denotes the average over all $d=(K^2-K)/2=46 971$ independent correlation coefficients of every correlation matrix $C(t)$. 

\begin{figure}[h] 
\centerline{
\includegraphics[width=1 \textwidth]{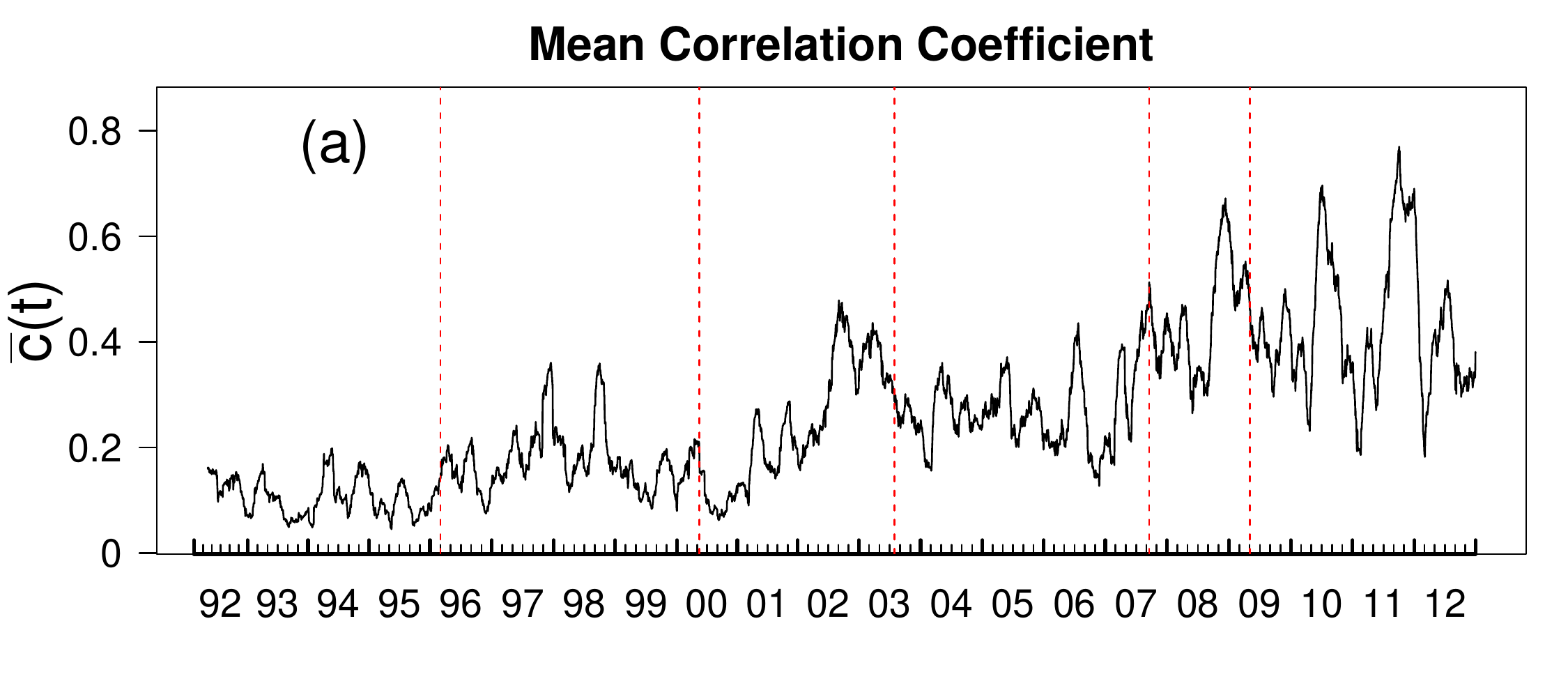}}\centerline{\includegraphics[width=1 \textwidth]{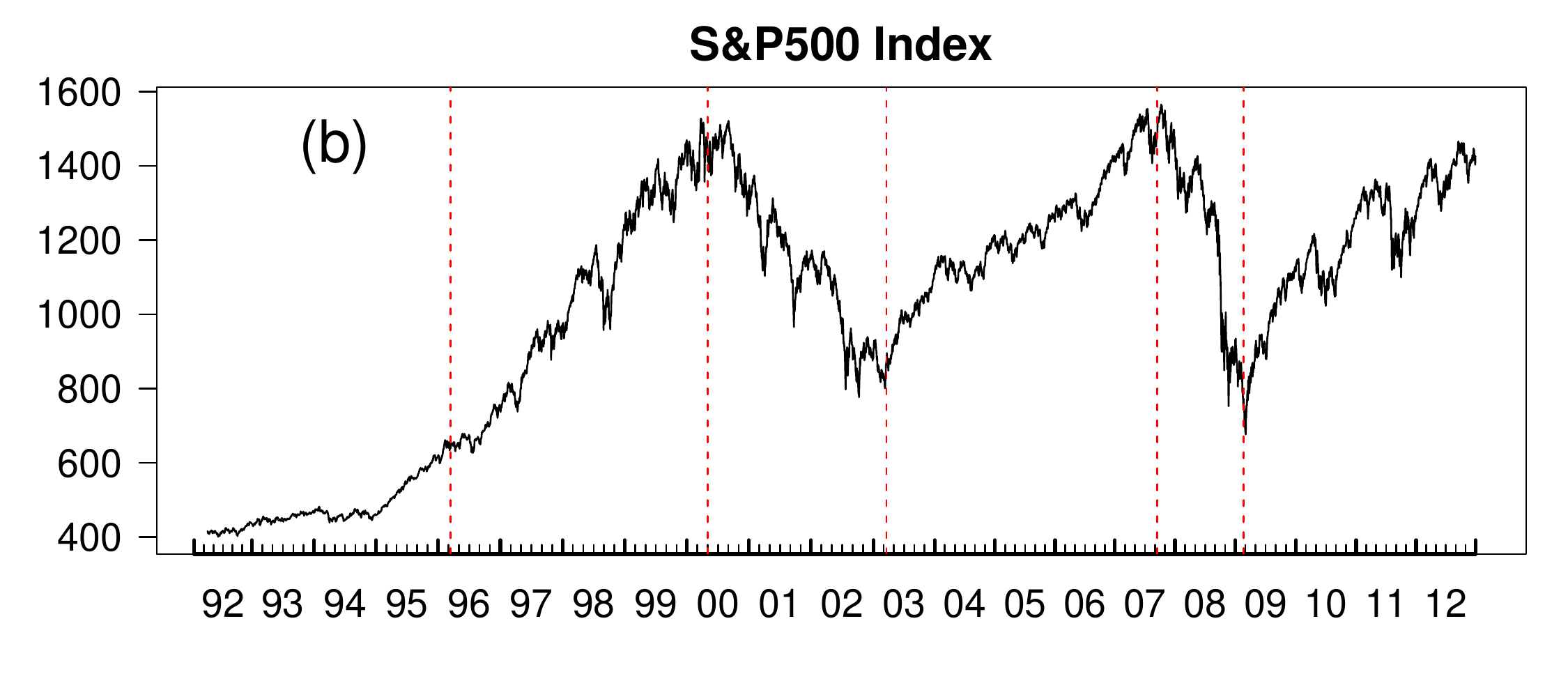}}
\caption{\label{fig:MeanCorr}
(a) Time evolution of the mean correlation coefficient $\bar{c}(t)$. (b) The S$\&$P500 Index for the same time period. Dashed lines highlight economically distinct time intervals as described in Section \ref{sec:periods}.}
\end{figure}

We recall the spectral decomposition  
\begin{equation} \label{eq:spectral}
C(t)=\underset{a=1}{\overset{T}{\sum}}\lambda_{a}(t)\vec{u}_{a}(t)\vec{u}_{a}^{\dagger}(t)=\lambda_{1}(t)\left(\vec{u}_{1}(t)\vec{u}_{1}^{\dagger}(t)+\underset{a=2}{\overset{T}{\sum}}\mathcal{O}\left(\frac{\lambda_{a}(t)}{\lambda_{1}(t)}\right)\right)
\end{equation} of the $K\times K$ correlation matrix $C(t)$ \cite{eigen,eigen2}. Here $\lambda_a(t)$ denotes the $a$th eigenvalue of $C(t)$, $\vec{u}_a(t)$ the corresponding normalized eigenvector and $\vec{u}_{a}^{\dagger}(t)$ its transpose. The rank of $C(t)$ is $T$ and therefore only the first $T$ eigenvalues are non-zero. For our data
the first and the largest eigenvalue $\lambda_{1}(t)=\lambda_\textrm{max}(t)$ is sufficiency larger than the other eigenvalues.  All components of $\vec{u}_1(t)$ are approximately equal to 0.05, while the components of the other $T-1$ eigenvectors spread around zero for every time $t$. Therefore $\vec{u}_1(t)$ corresponds to the dynamics of the whole market  as in Refs. \cite{eigen,eigen2}. Hence averaging over the correlation coefficients
\begin{equation}
\bar{c}(t)=\braket{C(t)}_{ij}\approx \kappa \lambda_\textrm{max}(t) 
\end{equation} we recover the largest eigenvalue. Here 
\begin{equation}
 \kappa =\braket{\vec{u}_{1}(t)\vec{u}_{1}^{\dagger}(t)}_{ij}\approx 229
\end{equation}
 is an empirical factor which appears due to the noise in the data. The time evolution of the largest eigenvalue is strongly correlated with the mean correlation coefficient $\bar{c}(t)$, the Pearson correlation is 0.998.  The quantities $\lambda_\textrm{max}(t)$ and $\bar{c}(t)$ share therefore the same dynamics. We will show in Sec. \ref{sec:pca} that $\bar{c}(t)$ has as much variability in the values as possible for our data.  Figure \ref{fig:MeanCorr} (a) shows the time evolution of $\bar{c}(t)$. We also present the time evolution of the S$\&$P500 Index in Fig. \ref{fig:MeanCorr} (b). 

\subsection{Geometric Approach: Principal Component Analysis} \label{sec:pca}
We identify each  correlation matrix $C(t)$ with a correlation vector
\begin{equation}
\vec{c}(t)=\left[\begin{array}{c}
c_{1}(t)\\
c_{2}(t)\\
\ldots\\
c_{d}(t)
\end{array}\right]
\end{equation}
 in the real $d$-dimensional Euclidian space $\mathbb{R}^d$. Here $c_i(t)$ is the $i$th component of $\vec{c}(t)$. We then apply the principal component analysis (Pearson \cite{pearson1901lap}, Hotelling \cite{Hotelling}) to quantify orthogonal and therefore uncorrelated one-dimensional subspaces in our time series $c_i(t)$, $i=1,...,d$.
 
The first principal component is defined as the line in $\mathbb{R}^d$ with the largest possible variance of the data values. The other principal components  are those with the largest data variance and orthogonal to the preceding components. The number of the principal components is smaller or equal to $d$. The principal components are spanned by the orthogonal eigenvectors $\hat{v}_i$, $i=1,..,d$ of the symmetric $d\times d$ covariance  matrix
\begin{equation} \label{eq:covariance}
W=A A^{\dagger}.
\end{equation} Here $A$ is the $d \times T$ data matrix with $d$ empirical times series $c_i(t)$ as rows and  $A^{\dagger}$ denotes its transpose. 

\begin{figure}[h] 
\centerline{
\includegraphics[width=0.5 \textwidth]{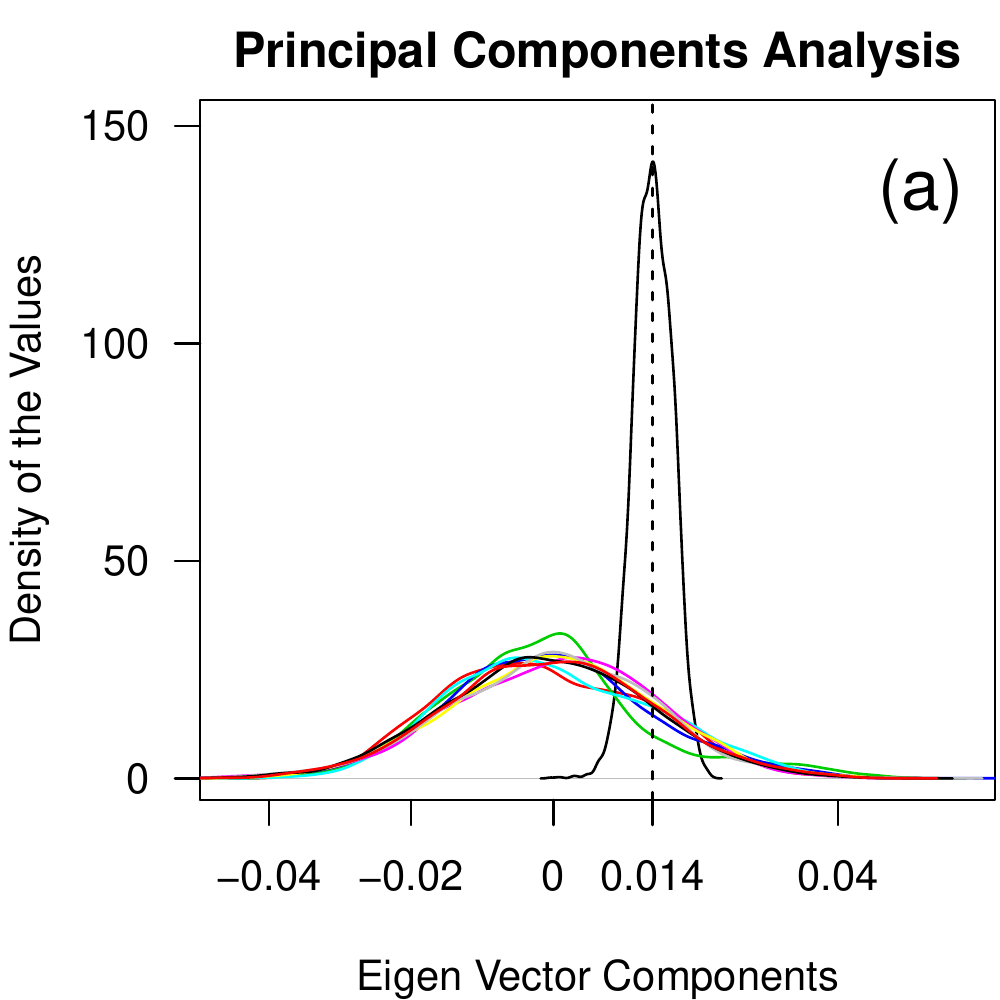}
\includegraphics[width=0.5 \textwidth]{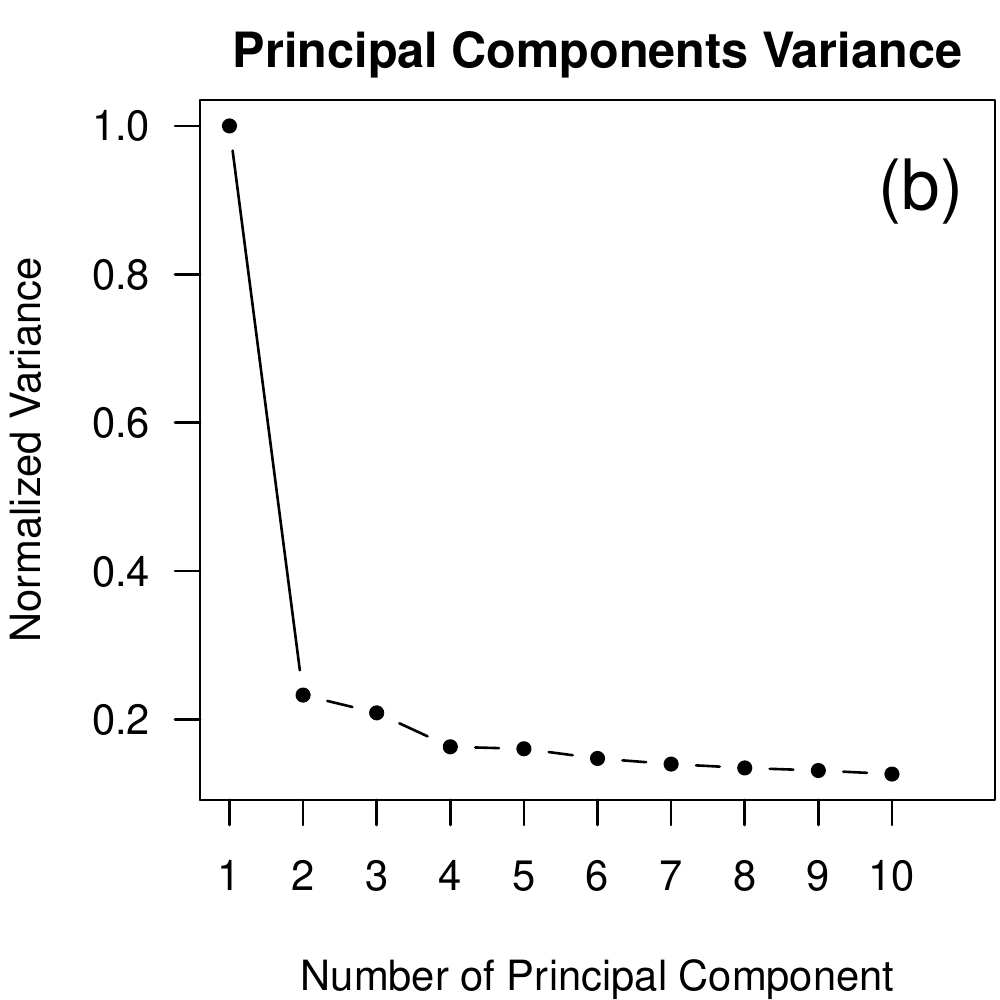}}
\caption{(a) The distribution of the first ten normalized principal components and (b) their variances normalized to the largest value. \label{fig:PCA}}
\end{figure}

The rank of $W$ is $\mbox{min}(d,T)$ and we can not apply the PCA to our full data so we applied the principal  component analysis (PCA) to randomly chosen 100 stocks ending up with  $d=(100^2-100)/2=4950$ time series of length $T=5169$. Fig.~\ref{fig:PCA} (a) shows the eigen vector components distribution for the first ten principal components.

 The components of the first normalized  eigen vector are concentrated around a constant value 0.014, while the values of the other nine are symmetrically distributed around zero.  Therefore the direction with the largest variance in data values is the subspace spanned by the vector
\begin{equation} \label{eq:pcaev}
\hat{v}_1\equiv \frac{1}{\sqrt{d}} \left[\begin{array}{c}
1\\
1\\
...\\
1\\
\end{array}\right] \approx \left[\begin{array}{c}
0.014\\
0.014\\
...\\
0.014\\
\end{array}\right].
\end{equation} The variance of data values for the first ten principal components are shown in Fig. \ref{fig:PCA} (b). The variance of the first principal component is much larger than the others. The correlation matrices $C(t)$ from our data set seen as vectors $\vec{c}(t) \in \mathbb{R}^d$ are thus distributed along $\hat{v}_1$. 
\begin{figure}[h] 
\centerline{
\includegraphics[width=0.5 \textwidth]{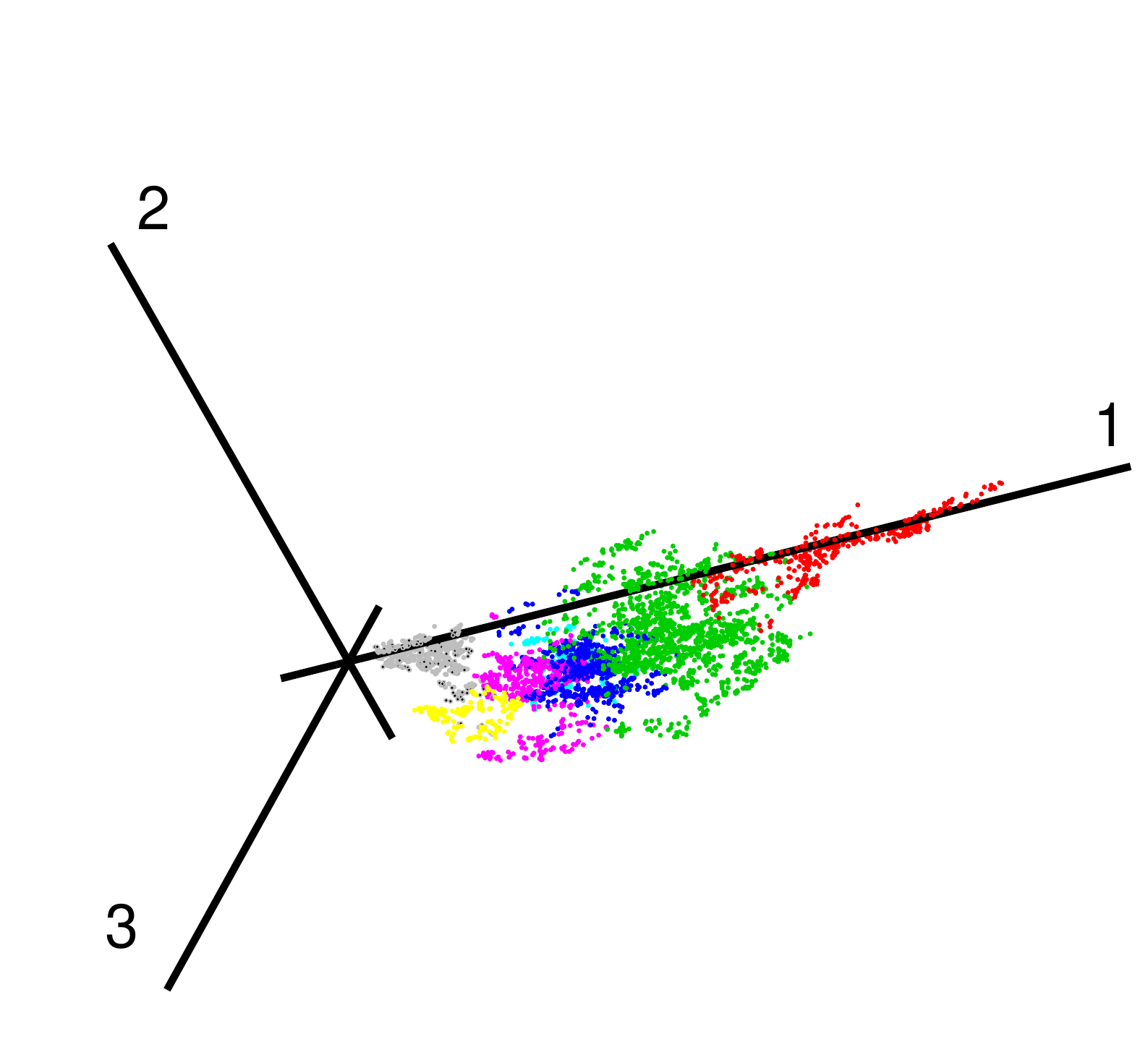}
\includegraphics[width=0.5 \textwidth]{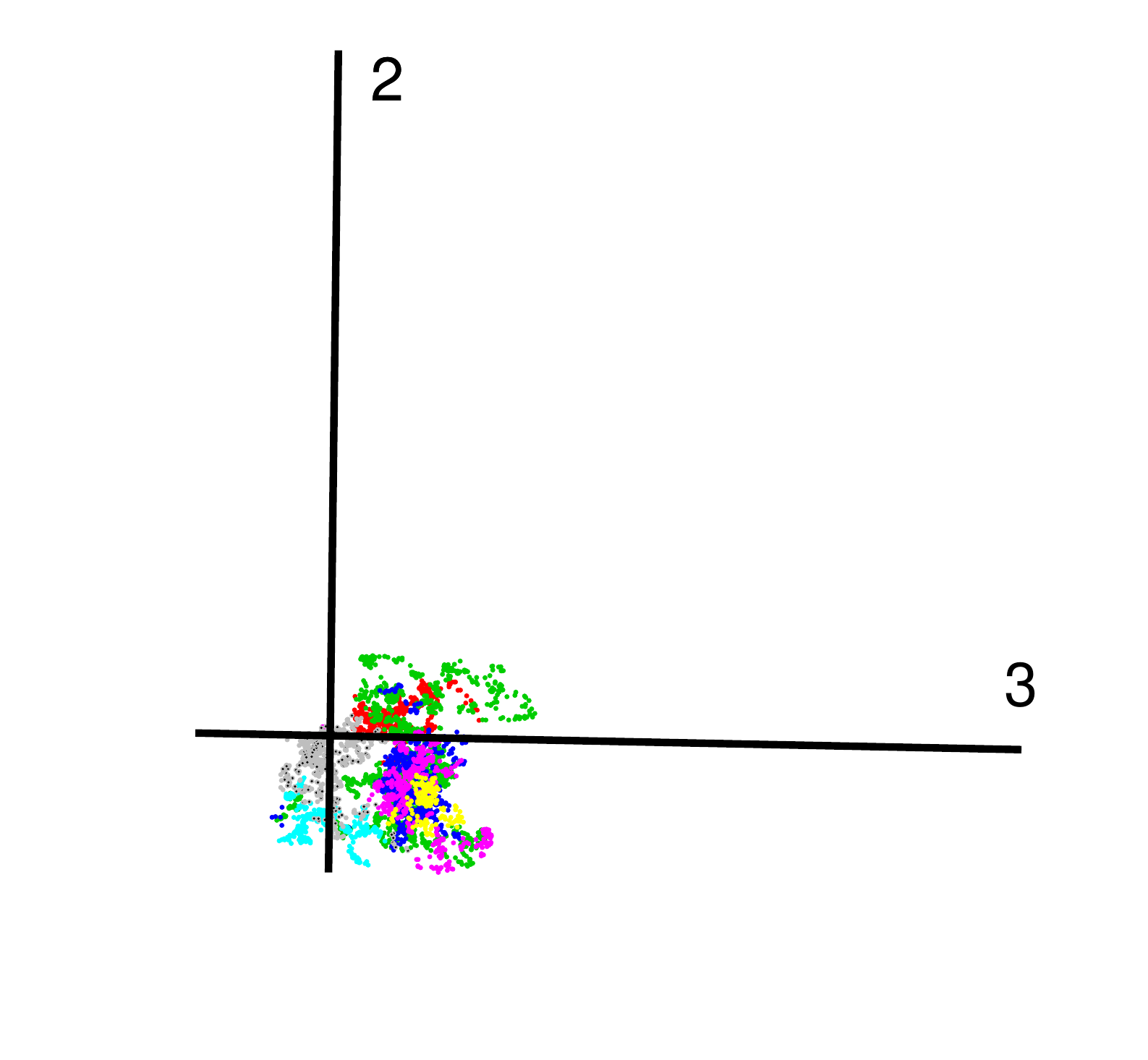}}
\caption{(Color online) Data set projected onto the first three pricipal components. Different colors highlight different  market states as explained in Sec.~\ref{fig:states}. \label{fig:PCA3d}}
\end{figure}
Figure~\ref{fig:PCA3d} shows the projection of our data onto the first three principal components in a scatter plot.  The distribution of the data points along the first principal component is dominating. The contribution of the correlation matrix $C(t)$ to the first principal component at time $t$ is given by the scalar product
\begin{equation} \label{eq:1stpc}
\left\langle \vec{c}(t), \hat{v}_1 \right\rangle=\frac{1}{\sqrt{d}} \underset{i=1}{\overset{d}{\sum}} c_i(t) = \bar{c}(t) \sqrt{d},
\end{equation} and turns out to be the mean correlation coefficient \eqref{eq:meancor} times the fixed number $\sqrt{d}$. The dynamics of the market is therefore dominated by the movement along $\hat{v}_1$ which is given by $\bar{c}(t)$. Eq.~\eqref{eq:1stpc} confirms the spectral analysis results discussed in Sec.~\ref{sec:quanti}. We note that spectral analysis of the correlation matrix $C(t)$ is the principal component analysis of the standardized returns 
\begin{equation} \label{eq:ret}
 \hat{r}_i(t)=\frac{r_i(t)-\braket{r_{i}}_T(t)}{\sigma^{(T)}_{i}(t)} .
\end{equation} treated as element of $\mathbb{R}^K$. Therefore the projection of \eqref{eq:ret} on the first principal component in $\mathbb{R}^K$ at time $t$ is equal to the non-weighted average of $\hat{r}_i(t)$.

\begin{figure}[h] 
\centerline{
\includegraphics[width=1 \textwidth]{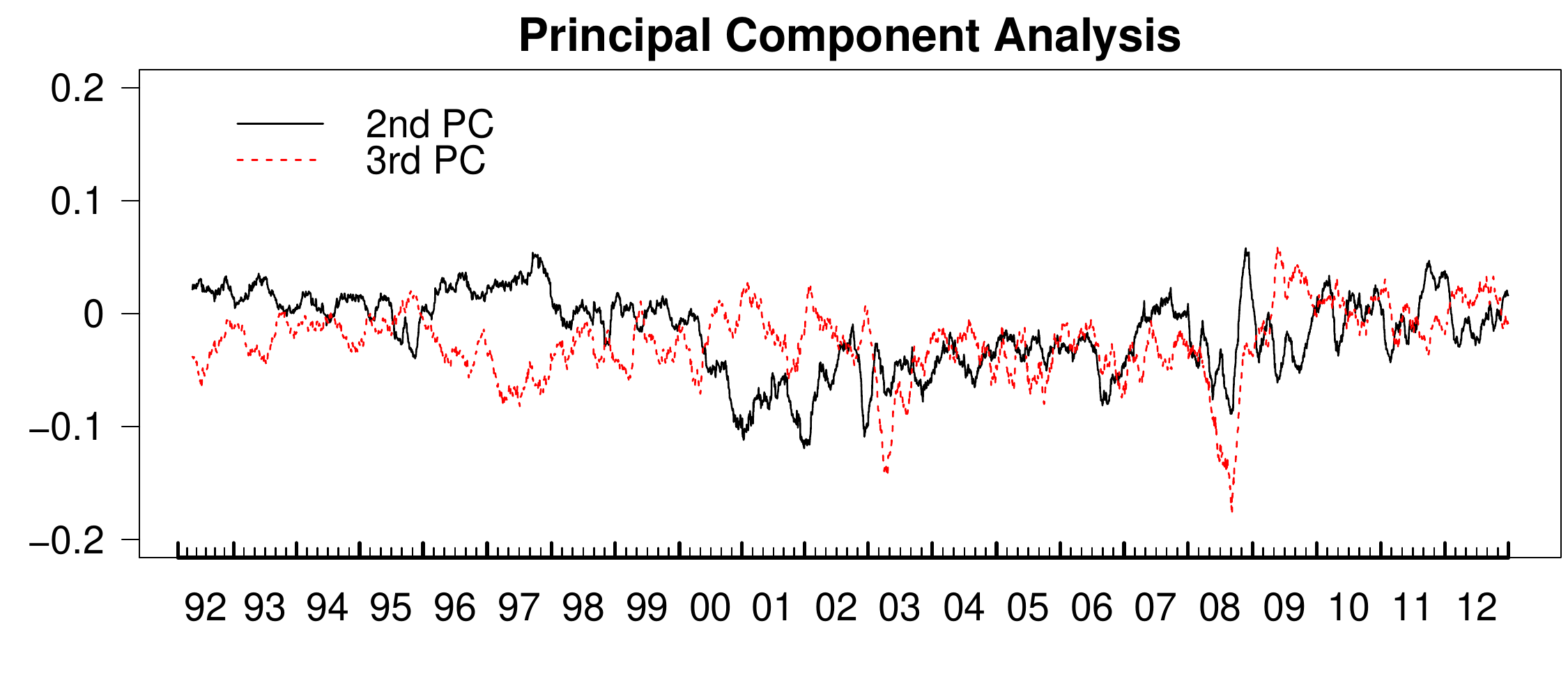}}
\caption{(Color online) Time evolution of the projections  $\left\langle \vec{c}(t), \hat{v}_2 \right\rangle$ (solid, black)  and $\left\langle \vec{c}(t), \hat{v}_3 \right\rangle$ (dashed, red) onto the second and third  principal components normalized by $\sqrt{d}$. \label{fig:PCATS}}
\end{figure}

The projections $\left\langle \vec{c}(t), \hat{v}_2 \right\rangle$ and $\left\langle \vec{c}(t), \hat{v}_3 \right\rangle$ describe system dynamics along the second and third principal component and are shown in Fig.~\ref{fig:PCATS}.

\section{Market States: Distinct Periods of the Market}  \label{sec:msperiod}
We cluster the data following Ref. \cite{ms} and identify the quasi-stationary states of the financial market which we present in Sec.~\ref{sec:ms}. We connect the characteristic states on the market to the known historical events in Sec.~\ref{sec:periods}. 

\subsection{Market States} \label{sec:ms}

\begin{figure}[h] 
\includegraphics[width=1 \textwidth]{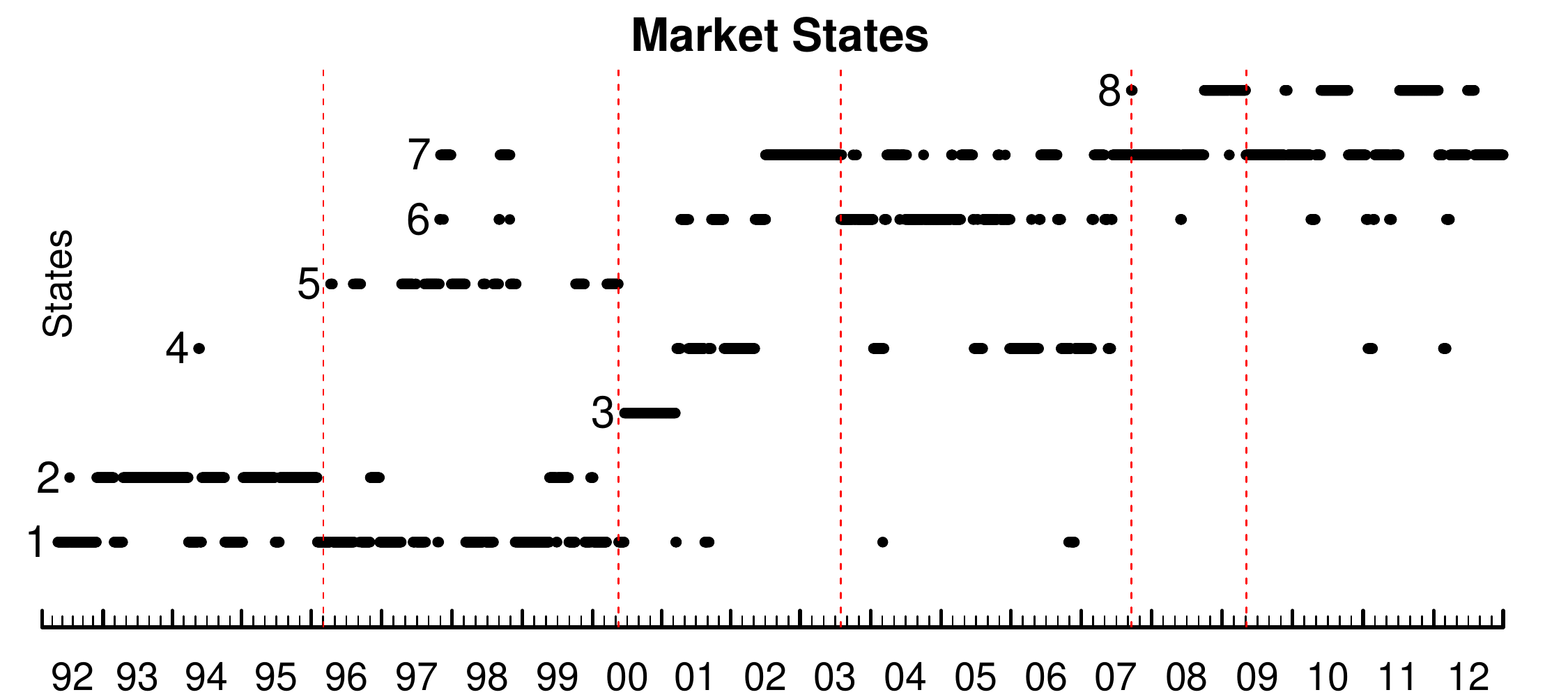}
\caption{\label{fig:states}
Time evolution of the market states. Dashed lines highlight economically distinct time intervals as described in Section \ref{sec:periods}}
\end{figure}

\begin{figure}[h] 
\centerline{\includegraphics[width=0.6 \textwidth]{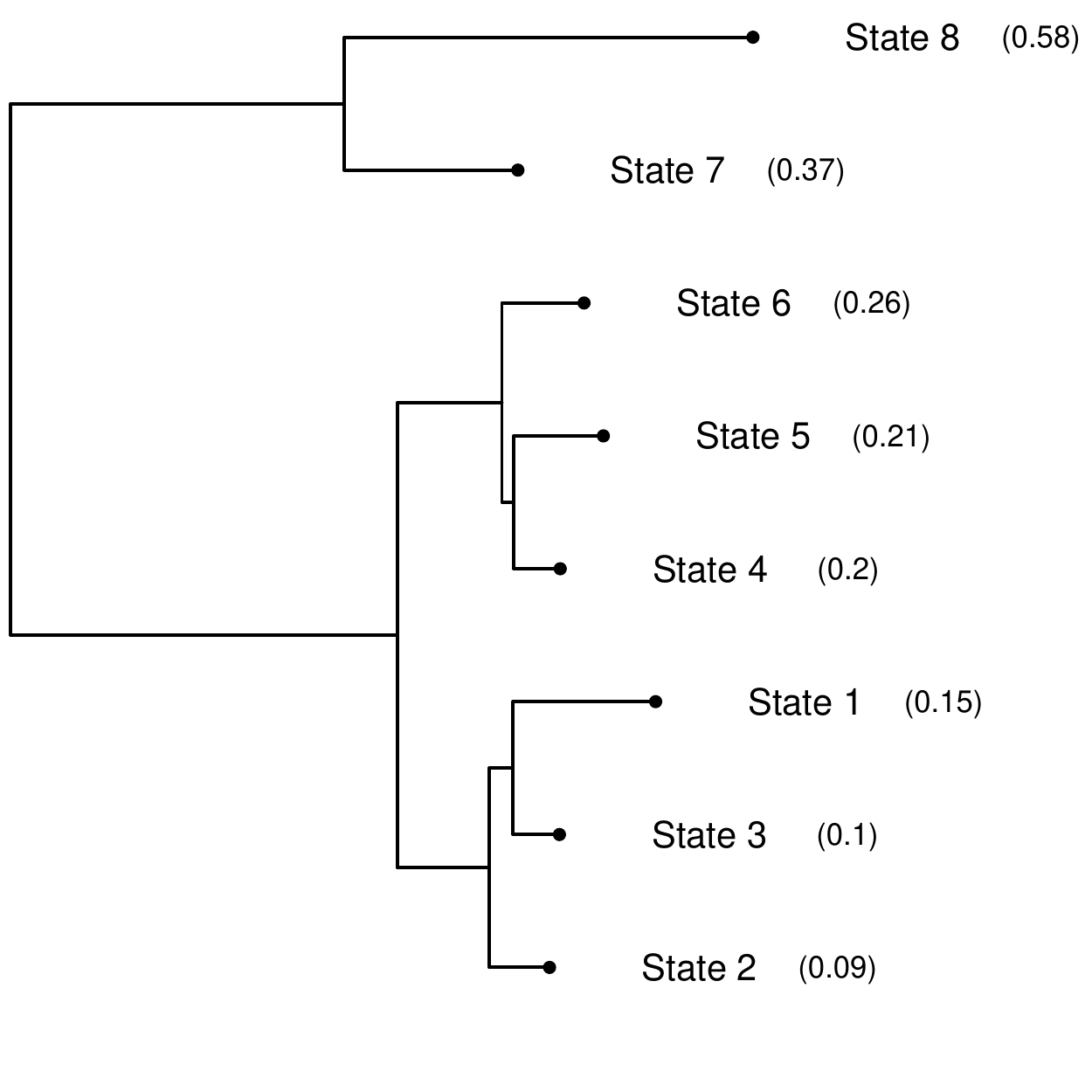}}
\caption{\label{fig:dendo}
Clustering tree of the market states clustering.  The mean value of $\bar{c}(t)$ within the states is given in parentheses.}
\end{figure}

In the previous section we showed that our data is spread along a few dominating subspaces in $\mathbb{R}^d$. To quantify the similarity between any two correlation matrices $C(t_a)$ and $C(t_b)$ we calculate the distance
\begin{equation}
D_{ab}=\parallel C(t_a)-C(t_b) \parallel = \parallel \vec{c}(t_a)-\vec{c}(t_b) \parallel
\end{equation} via the Euclidean norm on $\mathbb{R}^d$ normalized by $\sqrt{d}$.

As the next step we use the bisecting \textit{k}-means clustering algorithm \cite{Steinbach2000}. At the beginning of the clustering procedure all of the correlation matrices are considered as one cluster, which is then divided into two sub clusters using the $k$-means algorithm with $k = 2$. For each cluster $\alpha$ we then calculate its cluster center 
\begin{equation} 
\vec{\mu}_{\alpha} =\frac{1}{N_\alpha}\underset{t\in\alpha}{\sum}\vec{c}(t),
\end{equation} which is the mean correlation matrix in this cluster. Here  $N_\alpha$ denotes the number of the cluster elements and $t\in\alpha$ symbolically denotes all time $t$ for which $\vec{c}(t)$ is in the cluster $\alpha$. The separation procedure is repeated until the cluster  size
\begin{equation} 
R_{\alpha} = \frac{1}{N_\alpha}\underset{t\in\alpha}{\sum}\parallel \vec{c}(t)-\vec{\mu}_{\alpha} \parallel
\end{equation} is smaller than a given threshold for every cluster $\alpha$. We choose the mean distance to be smaller than 0.164 to achieve 8 clusters as in Ref.~\cite{ms}.
The market is said to be in a market state $\alpha$ at time $t$, if the corresponding correlation matrix $C(t)$, and hence the correlation vector $\vec{c}(t)$, is in the  cluster $\alpha$. 
The time evolution of the market states is shown in Fig.~\ref{fig:states}. In Figure~\ref{fig:dendo} the corresponding clustering tree is shown. The state occupied on the first day of our data is labeled by one. The remaining states are labeled according to the mean value of $\bar{c}(t)$ within the states as shown in Fig.~\ref{fig:dendo}. We group the states into three main classes. The market states one, two and three represent calm states. The states four, five and six are intermediate states. The states seven and eight are the turbulent states. The financial market evolves between these different states. New states form and existing states vanish in the course of time. For example the first four years are dominated by the states 1 and 2 in the last four years mainly the states 8 and 7 are occupied.

\subsection{Distinct Time Periods} \label{sec:periods}
We divide the entire time period into six dynamically and economically distinct intervals.
\begin{enumerate} 
\item Early 1992 to spring of 1996: in this rather calm period  $\bar{c}(t)$ varies between 0 and 0.2. The S$\&$P500 Index continuously grows with moderate volatility. The market mainly occupies the first and the second state.

\item From spring 1996 until spring 2000: the range that $\bar{c}(t)$ explores as well as the S$\&$P500 Index drastically increase. The volatility also becomes larger.  The increase of $\bar{c}(t)$ is explained by the appearance of strongly correlated industrial sectors during this period, especially the technology sector. The market state two almost disappears and the market jumps mainly between states five and one. We note that the fifth state appears only during this period. 

\item Spring 2000 to the second half of 2003: this period  fully covers the \textit{dot-com bubble}  and is known as a very turbulent time in financial markets due to the crisis. The S$\&$P500 Index drops continuously, losing about  half of its value. The mean correlation coefficient reaches its maximum at 0.48. At the beginning of the crises state 3 appears for about one year. This state appeared only once during the entire time period. In the second half the market is switching between states four and six and occupies state seven by the end of 2002. This period includes the market response to the 9/11 attacks.

\item From the second half of 2003 until fall of 2007: this period covers the four years period before the recent global financial crises up to the 1 year period before the collapse of \textit{Lehman Brothers}. As seen from  the S$\&$P500 Index in Fig. \ref{fig:MeanCorr}, the market seems to recover after the dot-com crisis but $\bar{c}(t)$ does not calm down and strongly fluctuates around  a mean value  0.28. The market is jumping between states four, six and seven. State six is occupied mainly during this interval.
 
\item From October 2007 until March 2009: this period covers the late-2000s financial crisis. The S$\&$P500 Index drops continuously and looses approximately half of its value. The mean correlation coefficient is peaked sharply at 0.67. The market is mainly in state seven and occupies the eighth state by the end of 2008.

\item March 2009 to end 2012:  the market seems  to slowly recover as the S$\&$P500 Index grows again. The growth interrupted by drastic drops.  This is reflected in high peaks of $\bar{c}(t)$, which accounts its maximum value 0.77 in the analyzed 21 years. The mean correlation coefficient does not relax to the values it had before the crisis.  The market is switching between states seven and eight and decays for short time into the states  four and six.
\end{enumerate}

\section{Stochastic Analysis}  \label{sec:stochAnal}
We describe the stochastic process used to model $\bar{c}(t)$  in Sec.~\ref{sec:sde}. In Sec.~\ref{sec:esti} we explain how the explicit model is extracted form the time series. We describe the stochastic analysis of the market states in Sec.~\ref{sec:msdyn}.
\subsection{Stochastic Processes} \label{sec:sde} 
We model $\bar{c}(t)$ as a stochastic process described by a Langevin equation 
\begin{equation} \label{eq:diffeq}
\frac{\mbox{d}}{\mbox{d}t}\bar{c}(t)=f(\bar{c},t) + g(\bar{c},t)\Gamma(t),
\end{equation}\textit{i.e.} a stochastic differential equation (SDE) for the variable $\bar{c}(t) \in \mathbb{R}$. Here $f$ is the deterministic part of \eqref{eq:diffeq} - the drift function and $g$ is the diffusion function, which defines the stochastic part of \eqref{eq:diffeq}. $\Gamma(t)$ is the $\delta$-correlated Gaussian white noise with $\langle \Gamma(t) \rangle=0$ and $\langle \Gamma(t_1)\Gamma(t_2) \rangle=\delta(t_1-t_2)$. We note that for the dimensionless  variable $\bar{c}(t)$ the drift function has a dimension of inverse time and the diffusion function has a dimension of inverse square root of time.

The solution of \eqref{eq:diffeq} is defined in terms of stochastic integrals, which depend on the choice of the discretization \cite{itovsstra,sussmann,irrevers}. Throughout this paper we use It$\hat{\text{o}}$'s choice (see It$\hat{\text{o}}$'s interpretation of SDEs \cite{itovsstra,ito1944}). The advantage of It$\hat{\text{o}}$'s definition is that the diffusion term $g$ is uncorrelated with the Gaussian white noise $\left<g(\bar{c},t)\Gamma(t)\right>=0$ \cite{itovsstra}. The drift and diffusion terms can  therefore be obtained as  conditional moments \cite{itovsstra,risken1996}
\begin{align}
f(c,t)&=\underset{\tau\rightarrow0}{\mbox{lim}}\frac{\left<\bar{c}(t+\tau)-\bar{c}(t)\right>}{\tau} \bigg|_{\bar{c}(t)=c}, \label{eq:KM1}\\ 
g^{2}(c,t)&=\underset{\tau\rightarrow0}{\mbox{lim}}\frac{\left<(\bar{c}(t+\tau)-\bar{c}(t))^{2}\right>}{\tau}\bigg|_{\bar{c}(t)=c} . \label{eq:KM2}
\end{align} Here $c$ denotes the value of the stochastic variable $\bar{c}(t)$ at which the value of the drift or the diffusion is evaluated. At this one instant we distinguish between $\bar{c}(t)$ and a particular numerical value $c$. The average in Eqs. \eqref{eq:KM1} and \eqref{eq:KM2} is performed over all realizations of $\bar{c}(t)$ for which the condition $\bar{c}(t) =c$ holds. These equations  express therefore the time derivative of the mean displacement and its square of $\bar{c}(t)$ at $c$.

Expressions \eref{eq:KM1} and \eqref{eq:KM2}  allow one to estimate the drift and diffusion directly from the empirical data as shown in Refs. \cite{Siegert1998,Friedrich2000-1} and sketched below, see Ref. \cite{Renner2001,Friedrich2000-2,Friedrich2011,Wosnitza2014228} for applications.  In the present work we model $\bar{c}(t)$ by an It$\hat{\text{o}}$ stochastic process and estimate the deterministic as well as the stochastic part of the corresponding SDE from the empirical time series. 

\subsection{Estimation of the Conditional Moments} \label{sec:esti}

For the estimation of the drift and the diffusion  directly from the data set we mainly follow Refs. \cite{Siegert1998,Renner2001,Friedrich2000-1,Friedrich2000-2}. Here, we briefly sketch the estimation procedure for the drift function, \textit{i.e.} the first conditional moment \eqref{eq:KM1}, as the estimation of the diffusion function \eqref{eq:KM2} works accordingly.  
We first introduce a new function
\begin{equation}
M_c(\tau)=\frac{\left<\bar{c}(t+\tau)-\bar{c}(t)\right>}{\tau} \bigg|_{\bar{c}(t)=c}, \label{eq:h}
\end{equation} for which the drift function 
\begin{equation}
f(c,t)=\underset{\tau\rightarrow0}{\mbox{lim}} M_{c}(\tau)
\end{equation} is obtained at $\tau=0$. We note that we dropped the time variable $t$ in the argument of $M$ in Eq. \eqref{eq:h} for brevity. For the estimation of $M_{c}(\tau)$ at fixed $c$ as a function of $\tau$  we  divide the time series $\bar{c}(t)$ into bins with equal number of data points. For every bin $I$ the function $M_{c}(\tau)$ is then estimated as
\begin{equation} \label{eq:hemp}
M_{\bar{c}_{I}}(\tau) =\frac{\left<\bar{c}(t+\tau)-\bar{c}(t)\right>}{\tau} \bigg|_{\bar{c}(t)\in I}.
\end{equation}
Here $\bar{c}_I$ is the mean value of $\bar{c}(t)$  in  bin $I$ and the average is performed over all data  in this bin. We note that for the empirical data this estimation can  only be done for  discrete values of $\tau=1,2,3...$. We then fit a second order polynomial in $\tau$ to the empirically estimated values of \eqref{eq:hemp}, extracting the desired value of the drift at $\bar{c}_I$ as the constant coefficient of the fitted function. The estimation of \eqref{eq:h} is only possible for the realized values of the empirical times series $\bar{c}(t)$.

Instead of analyzing the drift function \eqref{eq:KM1} itself, it is more convenient to consider the potential function 
\begin{equation} \label{eq:potential}
V(\bar{c},t)=-\intop^{\bar{c}}f(x,t) \mbox{d}x,
\end{equation} defined as the negative primitive integral of $f$. The minus sign is a convention. The dynamics of the system is encoded in the shape of $V(\bar{c},t)$: the local minima of the potential function correspond to the quasi-stable equilibria, or quasi-stable fixed points, around which the system oscillates. In contrast, local maxima correspond to unstable fixed points. We note that potential functions are defined up to an additive constant. For the dimensionless variable $\bar{c}(t)$ the dimension of the potential function is the inverse time.
\subsection{Market States Dynamics} \label{sec:msdyn}

To quantify the market dynamics while it is in a fixed market state $\alpha $ we restrict the  estimation of \eqref{eq:hemp} and evaluate only the data points
\begin{equation}
\{\bar{c}(t),\bar{c}(t+\tau) \:|\: t \in \alpha\} \label{eq:cond1}
\end{equation} for each state $\alpha$. We therefore consider only displacements along the first principal component within the market states. No state transitions are allowed. Potential functions estimated this way provide information about the stability of the market states and reveal the fixed points.

As we mentioned in Sec.~\ref{sec:ms} we group the states into the three main classes according to the hierarchical structure as shown in Fig.~\ref{fig:dendo}. We estimate the potential functions for each class $A$ evaluating only the data points
\begin{equation}
\{\bar{c}(t),\bar{c}(t+\tau) \:|\: t \in A\}. \label{eq:cond2}
\end{equation} Here $t \in A$ symbolically denotes all time points at which market is in a state of the class A.  For example the market might be in the state 1 at time $t$ and in the state 2 at time $t+\tau$, as these two clusters belong to the same class. We therefore consider only displacements within  $A$ and allow for state transitions between state of the same class.

\section{Results} \label{sec:results}
We show the estimated diffusion function \eqref{eq:KM2} in Sec.~\ref{sec:diffusion} and discuss the estimated potential function \eqref{eq:potential} in Sec.~\ref{sec:ExtPot}. In Section \ref{sec:dotcom} we take a closer look at the dot-com bubble. A detailed study of the market states dynamics is presented in Sec. \ref{sec:msdyn}.

\subsection{Diffusion Term} \label{sec:diffusion}
\begin{figure}[h] 
\centerline{\includegraphics[width=1 \textwidth]{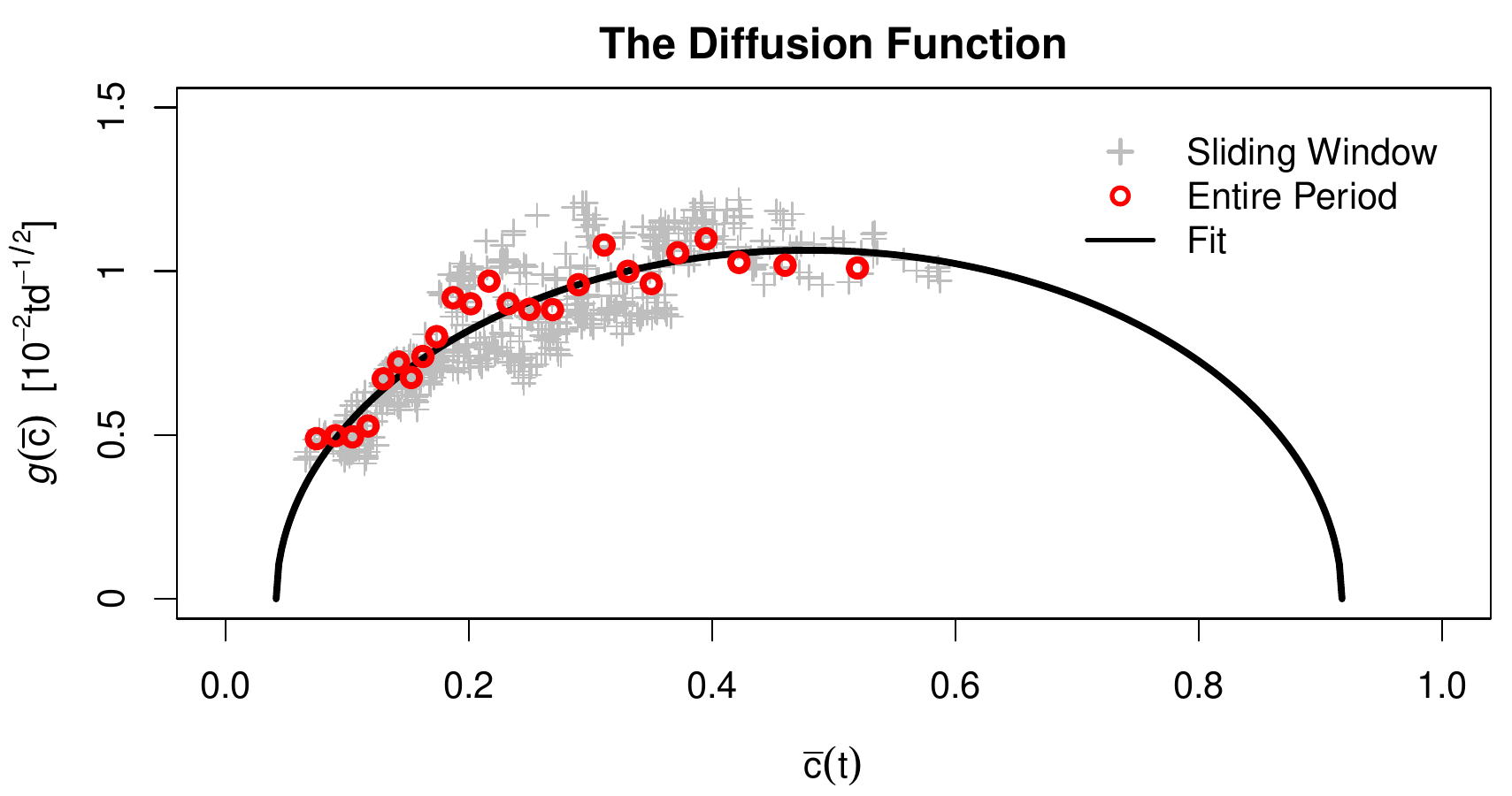}}
\caption{The diffusion function estimated on the sliding window (crosses (+)). The circles (o) show the estimated diffusion function on the entire time period at once. The solid curve shows the fitted function \eqref{eq:fitg}. We only use values estimated on the sliding windows for the fit. \label{fig:g}}
\end{figure}

To quantify the time dependency of the diffusion function $g(\bar{c},t)$ we estimated the  second conditional moment \eqref{eq:KM2} on a time window of four trading years (1008 trading days) which is moved in steps of two trading months (42 trading days). All together we obtain 100 estimates for $g(\bar{c},t)$ which we present in Fig.~\ref{fig:g}. As we explained in Sec.~\ref{sec:esti}, the estimation is only possible for the realized values of $\bar{c}(t)$. 
We therefore put all estimated values in a single diagram. We then fit the estimated values by the time-independent function
\begin{equation} \label{eq:fitg}
g(\bar{c})=\lambda \sqrt{ (\bar{c}-c_\textrm{min})(c_\textrm{max}-\bar{c})},  
\end{equation} which fits our data well, see Fig.~\ref{fig:g}. The diffusion function \eqref{eq:fitg} is widely used to model the stochastic correlation \cite{Ball:2000aa,Burtschell05beyondthe,corr,emmer,JunMa2009}, as it limits the values of the correlation to the range $[c_\textrm{min},c_\textrm{max}]$.  From the estimated parameters
\begin{eqnarray}
\lambda &=& 0.0245  \mbox{ td}^{-1/2}, \label{eq:lambda}\\
c_\textrm{min} &=& 0.042, \\
c_\textrm{max} &=& 0.918,
\end{eqnarray} we obtain the characteristic time scale of the system
\begin{equation}
t_0=\frac{1}{\lambda^2} = 1666 \mbox{ trading days},
\end{equation} which turns out to be approximately one third of the analyzed period.  For consistency we estimate \eqref{eq:KM2} for the entire time series $\bar{c}(t)$ at once, as shown in Fig.~\ref{fig:g}. We note that we fitted \eqref{eq:fitg} only to the data obtained on the sliding window.

\subsection{Time Evolution of the Potential Functions in the Entire Time Period} \label{sec:ExtPot}

\begin{figure}[t] 
\centerline{\includegraphics[width=.6 \textwidth]{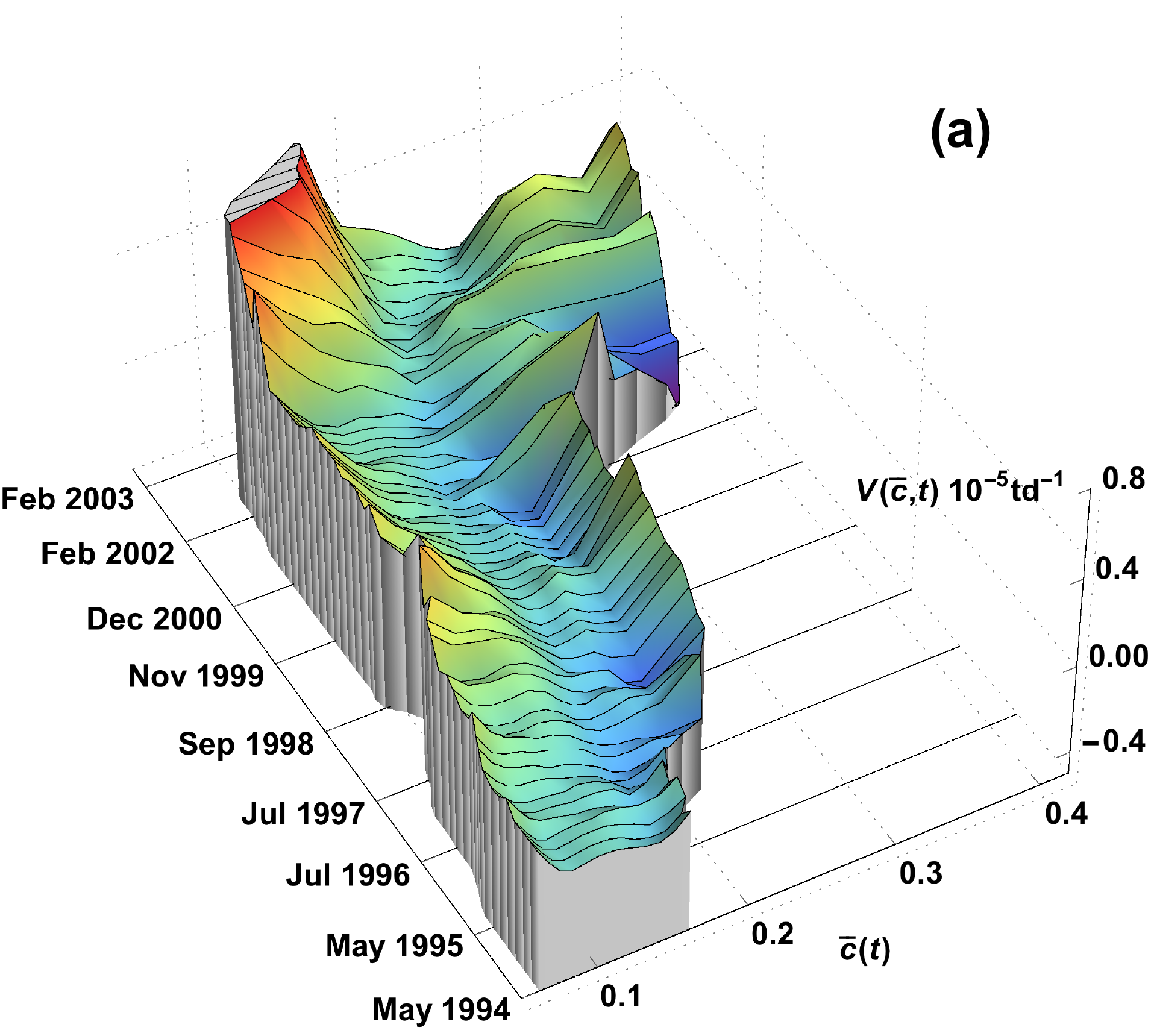}
\includegraphics[width=.6\textwidth]{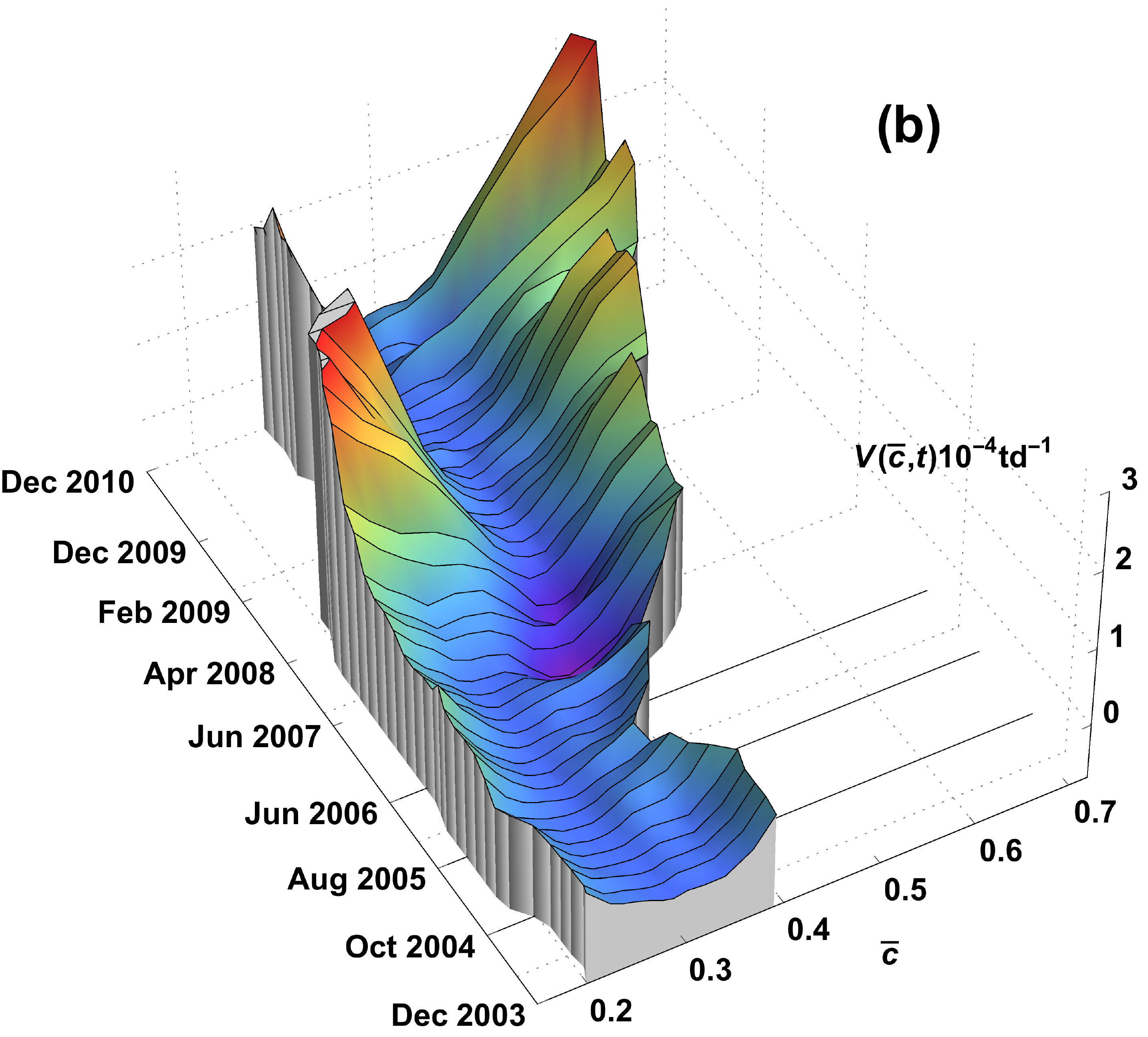}}
\caption{(Color online) Time evolution of the potential function \eqref{eq:potential} from early 1992 to the end of 2004 (a) and from 2002 to the end of 2012 (b) estimated on a time window of four trading years which is moved in steps of two trading months. The dates mark the time points in the middle of the estimation time windows. Representation is according to Eq. \eqref{eq:Vconstant}. \label{fig:potentil}}
\end{figure}

To quantify the time dependence of the drift function $f(\bar{c},t)$ we estimate the first conditional moment \eqref{eq:KM1} on a time window of four trading years (1008 trading days) which is moved in steps of two trading months (42 trading days). All together we obtain 100 estimates for $f(\bar{c},t)$. We then calculate the potential functions \eqref{eq:potential} which are presented in Fig.~\ref{fig:potentil}~(a)-(b). The dates mark the time points in the middle of the estimation time windows. In contrast to the diffusion function, the drift function turns out to be time-dependent. Therefore it is difficult to graphically present many curves in a single diagram, as the potential function \eqref{eq:potential} is defined up to an additive constant. To work around this problem we set 
\begin{equation} \label{eq:Vconstant}
V(\bar{c}_0,t)=0,
\end{equation} where $\bar{c}_0$ denotes the value at which $V(\bar{c},t)$  has its minimum in the first half of its values. In this representation the deeper a potential function is, the higher are the boundaries. 

Figure~\ref{fig:potentil}~(a) shows the results from early 1992 to the end of 2004. The distinct time periods described in Sec.~\ref{sec:periods} are clearly recognizable in the shape of the potential function. It is flat and approximately constant at the beginning. It gets deeper in the middle of the period during the turbulent time in 1997-98. The two local minima show instabilities on the market. By the end of the period the dot-com crises is reflected in the shape of $V(\bar{c},t)$. Its boundaries get higher and it has many deep minima at high values of $\bar{c}(t)$.

Figure~\ref{fig:potentil}~(b) shows the results from early 2002 to the end of 2012. Similar to the previous case, $V(\bar{c},t)$ is flat and constant during the relatively calm period in the early 2000s. It changes its shape drastically in the second half of the 2007 and gets a deep local minimum around $\bar{c}(t) \approx 0.4$. The boundaries get very high during the late-2000s financial crisis. We note that $V(\bar{c},t)$ does not become flat after 2010.

We showed that $\bar{c}(t)$ is described by a stochastic process \eqref{eq:diffeq} with a time-independent diffusion term and a time-dependent drift function. In  Sec.~\ref{sec:datanal} we showed that the mean correlation coefficient is the dominating variable of the collective market dynamics. The non-stationarity  of the potential function is therefore explained by deterministic changes in the collective correlation structure on the market.

\subsection{Zooming into the Dot-com Bubble} \label{sec:dotcom}

\begin{figure}[h] 
\centerline{\includegraphics[width=.7 \textwidth]{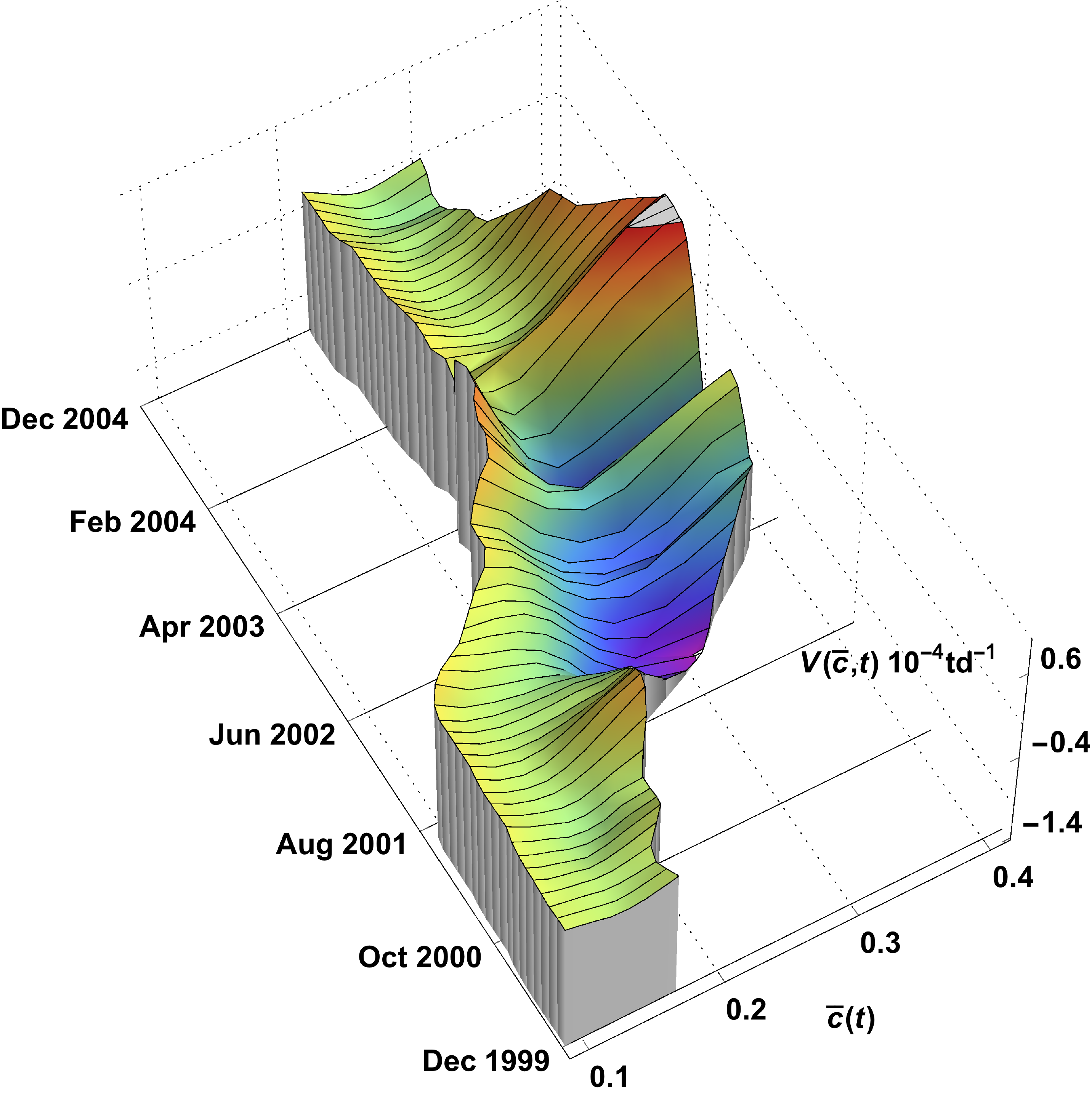}}
\caption{\label{fig:dotcom}
 (Color online) Time evolution of the potential function \eqref{eq:potential} in the period from early 1999 to early 2006, which covers the dot-com bubble. The estimation is done on a time window of two trading years which is moved in steps of one trading month. The dates mark the time points in the middle of the estimation time windows. Representation is according to Eq. \eqref{eq:Vconstant}. \label{fig:dotcom}}
\end{figure}

In the previous section we showed that the market evolves in time, switching back and forth between different market states.  As an example of a state transition we estimate $V(\bar{c},t)$ in the period from early 1999 to early 2006. The interval covers the dot-com bubble. To achieve higher time resolution  we  perform the estimation on a time window of two trading years (512 trading days), sliding it in steps of one trading month (21 trading days). Figure \ref{fig:dotcom} shows the time evolution of the estimated potential function. It is flat at the beginning where the market is mainly in the states 1 and 2, see Fig.~\ref{fig:states}. During the crisis the values of $\bar{c}(t)$ increase. Therefore the estimated potential function moves along the $\bar{c}$ axis. A deep minimum builds up and the boundaries get higher. The market jumps through the states 3, 4 and 6, ending up in state 7 by the end of 2002. By the end of 2003 the market settles into state 6 with only short jumps into the states 4 and 1. The potential function becomes constant but has changed its shape compared to the pre-crisis period. The market therefore jumps from a stable state to a turbulent state and then down to another stable state.

\subsection{Market States Dynamics: Stability, Hierarchy and State Transition} \label{sec:msdyn}

\begin{figure}[h] 
\centerline{\includegraphics[width=1.2 \textwidth]{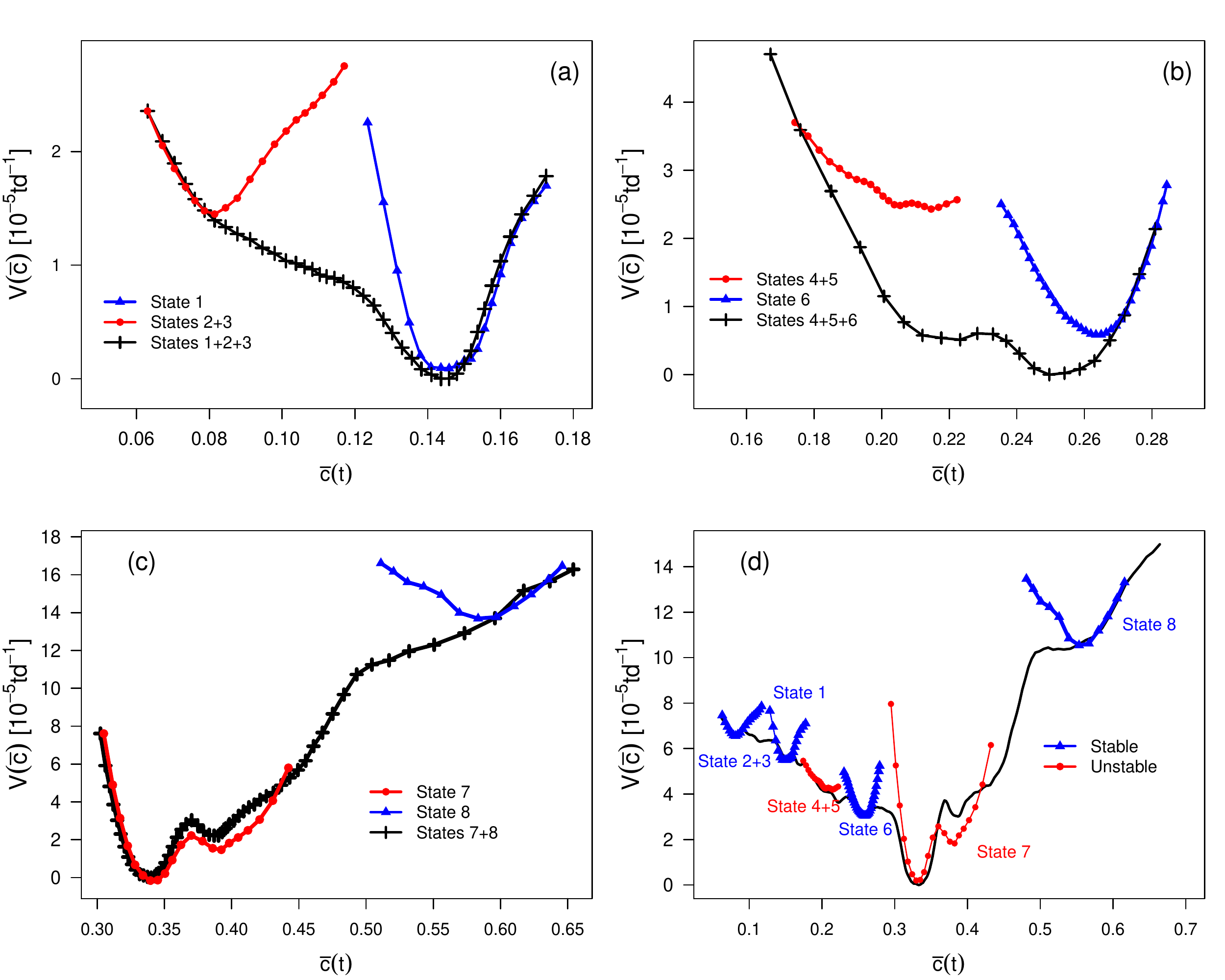}}
\caption{\label{fig:statePot}
(a)-(c) The potential function \eqref{eq:potential} of the displacements within the states \eqref{eq:cond1}  (filled circles and triangles). The potential function of displacement within the three groups \eqref{eq:cond2} is represented by the envelope line with crosses (\textbf{+}). (d) The overall potential landscape $V(\bar{c})$ estimated on the entire time series $\bar{c}(t)$ at once.}
\end{figure}

\begin{figure}[h] 
\centerline{\includegraphics[width=1 \textwidth]{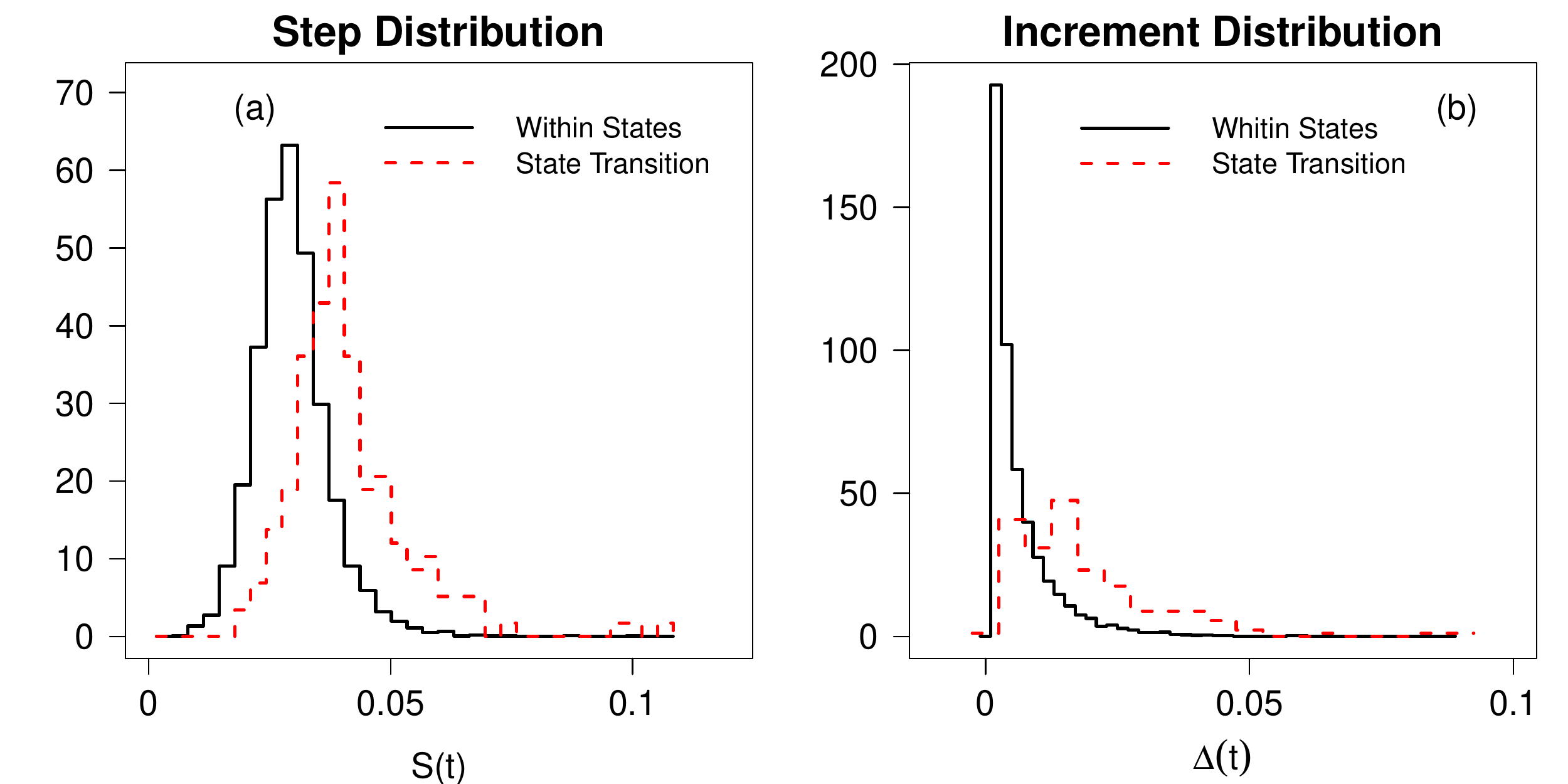}}
\caption{\label{fig:steps} (Color online)
Empirical histograms of (a) the steps \eqref{eq:steps} and (b) the increments \eqref{eq:increment} within states (black, solid) and during state transitions (red, dashed).}
\end{figure}

In the previous sections we showed that the mean correlation coefficient is described by a stochastic process \eqref{eq:diffeq} with the time-independent diffusion function \eqref{eq:fitg} and the time-dependent drift function. Especially, calm and turbulent periods can be distinguished by the shape of $V(\bar{c},t)$. To quantify the market dynamics in a given market state we estimate the potential function for the data points \eqref{eq:cond1}. Thus, we only accounted for displacements within a fixed market state.

The time series for the states 3, 4 and 5 are too short for the estimation of \eqref{eq:KM1}, so we combined the time series of the states 2 and 3 together as well as 4 and 5. We denote the resulting states by $2+3$ and $4+5$ respectively. As shown in Fig.~\ref{fig:dendo},  these pairs consist of states of the same class. Figure \ref{fig:statePot}~(a)-(c) shows the resulting potential functions for each market state. 

Potential  functions provide information about the stability of market states. This notion of the stability is not due to the time which the market spends occupying a certain state, but is given by the dynamics of the market. States 1, 2+3, 6 and 8 are stable states, as their potential functions have a single deep minimum and therefore a clearly defined fixed point. State 8 mainly appears during the latest financial crisis and represents a strong collective correlation on the market. In contrast state 7 is very unstable. Not only has its potential function two local minima, but it is also the deepest one. The correlation structure is non-stationary within the market state 7. The combined state 4+5 has a half-open potential function. States 4 and 5 are intermediate states between calm and turbulent periods, see Fig.~\ref{fig:states}. We note that within stable states $\bar{c}(t)$ is described by SDE \eqref{eq:diffeq} with the diffusion function \eqref{eq:fitg} and a linear drift function.

In Section \ref{sec:ms} we grouped the market states into three classes according to the clustering tree, see Fig.~\ref{fig:dendo}. Not all of the market states appear simultaneously  in a given time interval, as shown in Fig.~\ref{fig:states}. The first four years of the analyzed time period are dominated by the states 1 and 2, which belong to the first class.  In the last four years basically only the states 7 and 8 appear, which build the third class. To quantify the hierarchical structure of the states we estimate $V(\bar{c},t)$ for the points \eqref{eq:cond2}. We therefore account for displacements within the classes including state transitions. The resulting potential functions for the three classes are shown in Fig.~\ref{fig:statePot} (a)-(c). These curves envelope the potential functions of the market states of the corresponding class.

Similar to the envelopes we estimated $V(\bar{c})$ on the entire time period at once, as shown in Fig.~\ref{fig:statePot}~(d). The potential function of each market state has a distinct position along the first principal component, i.e. a distinct value of $\bar{c}(t)$. We therefore conclude that, while the market is in a given (stable) state, the mean correlation coefficient fluctuates around a mean value, which is defined by the minimum of the potential function, see Figs.~\ref{fig:statePot} and \ref{fig:dendo}. As we showed in Sec.~\ref{sec:pca}, the movement along the first principal component is given by the time evolution of $\bar{c}(t)$. Hence the market dynamics within a fixed state is given by the movement along the second and higher principal components, see Figs.~\ref{fig:PCA3d}~and~\ref{fig:PCATS}. Large changes of the mean correlation coefficient yield state transitions. The market is therefore "hopping" from state to state in the potential landscape, which is shown in Fig.~\ref{fig:statePot}~(d).  For consistency  we calculate the daily steps
\begin{equation} \label{eq:steps}
S(t)=\parallel \vec{c}(t+1)-\vec{c}(t) \parallel
\end{equation} of the market and the absolute increments
\begin{equation} \label{eq:increment}
\Delta (t) = \mid \bar{c}(t+1)-\bar{c}(t) \mid.
\end{equation}  of $\bar{c}(t)$. Figure \ref{fig:steps}~(a)-(b) shows the distribution of the steps \eqref{eq:steps} and the increments \eqref{eq:increment} within market states compared to the jumps during a state transition. Both the steps and especially  the increments are on average larger during state transitions that within states as we claimed.

\section{Conclusion}

The combination of geometric data analysis and stochastic methods sheds new light on the collective dynamics of complex systems. We applied these techniques to stock market data and evaluated the correlation structure on a sliding time window for a  period of 21 years. The collective market dynamics in terms of the principal components is given by the average correlation coefficient. We extracted the underlying stochastic process which turns out to have a time-independent stochastic term and a time-dependent deterministic term. The latter is represented graphically as a potential landscape and provides information on stability and system fixed points. We established the connection between distinct historical periods on the market and the time evolution of the potential function.  The non-stationary market dynamics can be attributed to changes in the deterministic part of the collective market dynamcis. 
We identified quasi-stationary states of the market following Ref.~\cite{ms} and  distinguished three main classes of market dynamcis: Calm, intermediate and turbulent states. To quantify the market states dynamics we estimated the potential functions, accounting only for displacements within a fixed state. 
In a given state the average correlation fluctuates around a distinct mean value, which defines a fixed point. 
The market dynamics within a market state is given by the movement along higher principal components. 
State transitions are reflected in large changes of the average correlation and correspond to the hopping in the potential landscape. 
Our results are consistent with the random matrix approach of Ref.~\cite{zomms} and contribute to a better overall understanding of market dynamics. 
While we highlighted the application to financial data in this paper, our approach should prove useful for the study of any quasi-stationary complex system.

\section{References}


\begin{thebibliography}{10}
\expandafter\ifx\csname url\endcsname\relax
  \def\url#1{{\tt #1}}\fi
\providecommand{\eprint}[2][]{\url{#2}}

\bibitem{Friedrich2011}
Friedrich R, Peinke J, Sahimi M and Tabar M~R~R 2011 {\em Physics
  Reports\/} {\bf 506} 87--162

\bibitem{nicolo}
Musmeci N, Aste T and Di~Matteo T 2015 {\em Journal of Network Theory in
  Finance\/} {\bf 1} 1--22

\bibitem{assettrees}
Onnela J~P, Chakraborti A, Kaski K, Kert{\'e}sz J and Kanto A 2003 {\em Physica
  Scripta\/} {\bf 2003} 48

\bibitem{Mizuno2006336}
Mizuno T, Takayasu H and Takayasu M 2006 {\em Physica A: Statistical Mechanics
  and its Applications\/} {\bf 364} 336 -- 342

\bibitem{rosario2}
Bonanno G, Caldarelli G, Lillo F, Miccich{\'e} S, Vandewalle N and Mantegna R~N
  2004 {\em The European Physical Journal B - Condensed Matter and Complex
  Systems\/} {\bf 38} 363--371

\bibitem{Tumminello201040}
Tumminello M, Lillo F and Mantegna R~N 2010 {\em Journal of Economic Behavior
  and Organization\/} {\bf 75} 40 -- 58

\bibitem{Tumminello26072005}
Tumminello M, Aste T, Di~Matteo T and Mantegna R~N 2005 {\em PNAS\/} {\bf 102}
  10421--10426

\bibitem{Pozzi:2013aa}
Pozzi F, Di~Matteo T and Aste T 2013 {\em Sci. Rep.\/} {\bf 3}

\bibitem{rosario}
Mantegna R~N 1999 {\em The European Physical Journal B - Condensed Matter and
  Complex Systems\/} {\bf 11} 193--197

\bibitem{ms}
M{\"u}nnix M~C, Shimada T, Sch{\"a}fer R, Leyvraz F, Seligman T~H, Guhr T and
  Stanley H~E 2012 {\em Sci. Rep.\/} {\bf 2}

\bibitem{kmeans}
Lloyd S 1982 {\em Information Theory, IEEE Transactions on\/} {\bf 28} 129--137

\bibitem{Steinbach2000}
Steinbach M, Karypis G and Kumar V 2000 A comparison of document clustering
  techniques {\em Sixth ACM SIGKDD International Conference on Knowledge
  Discovery and Data Mining\/} (Boston)

\bibitem{dbht}
Song W~M, Di~Matteo T and Aste T 2012 {\em PLoS ONE\/} {\bf 7}

\bibitem{Hotelling}
Hotelling H 1933 {\em Journal of Educational Psychology,\/} {\bf 24} 417--441,
  498--520

\bibitem{pearson1901lap}
Pearson K 1901 {\em Philosophical Magazine\/} {\bf 2} 559--572

\bibitem{NLDR}
Lee J~A and Verleysen M 2007 {\em Nonlinear Dimensionality Reduction\/}
  (Springer)

\bibitem{eigen}
Laloux L, Cizeau P, Potters M and Bouchaud J~P 2000 {\em International Journal
  of Theoretical and Applied Finance\/} {\bf 03} 391--397

\bibitem{eigen2}
Plerou V, Gopikrishnan P, Rosenow B, Amaral L~A~N, Guhr T and Stanley H~E 2002
  {\em Phys. Rev. E\/} {\bf 65}(6) 066126

\bibitem{Friedrich2000-1}
Friedrich R, Siegert S, Peinke J, L{\"u}ck S, Siefert M, Lindemann M, Raethjen
  J, Deuschl G and Pfister G 2000 {\em Physics Letters A\/} {\bf 271} 217 --
  222

\bibitem{Renner2001}
Renner C, Peinke J and Friedrich R 2001 {\em Physica A: Statistical Mechanics
  and its Applications\/} {\bf 298} 499 -- 520

\bibitem{philip}
Rinn P, Stepanov Y, Peinke J, Guhr T and Sch{\"a}fer R 2015 {\em
  arXiv:1502.07522\/}

\bibitem{PhysRevE.84.031103}
Vasconcelos V~V, Raischel F, Haase M, Peinke J, W\"achter M, Lind P~G and
  Kleinhans D 2011 {\em Phys. Rev. E\/} {\bf 84}(3) 031103

\bibitem{PhysRevE.61.R4691}
Hutt A, Svens\'en M, Kruggel F and Friedrich R 2000 {\em Phys. Rev. E\/} {\bf
  61}(5) R4691--R4693

\bibitem{locnorm}
Sch{\"a}fer R and Guhr T 2010 {\em Physica A: Statistical Mechanics and its
  Applications\/} {\bf 389} 3856 -- 3865

\bibitem{itovsstra}
van Kampen N 1981 {\em Journal of Statistical Physics\/} {\bf 24} 175--187

\bibitem{sussmann}
Sussmann H~J 1978 {\em The Annals of Probability\/} {\bf 6} 19--41

\bibitem{irrevers}
J~Dunkel S~Hilbert P~H 2006 {\em (In German) Irreversible Prozesse und
  Selbstorganisation\/} (Logos Verlag Berlin) chap Langevin-Gleichungen mit
  nichtlinearer Reibung, pp 11 -- 21

\bibitem{ito1944}
It{\^o} K 1944 {\em Proceedings of the Imperial Academy\/} {\bf 20} 519--524

\bibitem{risken1996}
Risken H 1996 {\em The Fokker-Planck Equation: Methods of Solution and
  Applications\/} Lecture Notes in Mathematics (Springer Berlin Heidelberg)

\bibitem{Siegert1998}
Siegert S, Friedrich R and Peinke J 1998 {\em Physics Letters A\/} {\bf 243}
  275 -- 280

\bibitem{Friedrich2000-2}
Friedrich R, Peinke J and Renner C 2000 {\em Phys. Rev. Lett.\/} {\bf 84}(22)
  5224--5227

\bibitem{Wosnitza2014228}
Wosnitza J~H and Leker J 2014 {\em Physica A: Statistical Mechanics and its
  Applications\/} {\bf 401} 228 -- 250

\bibitem{Ball:2000aa}
Ball C~A and Torous W~N 2000 {\em Journal of Empirical Finance\/} {\bf 7}
  373--388

\bibitem{Burtschell05beyondthe}
Burtschell X, Gregory J and Laurent J~P 2005 {\em Journal of Credit Risk\/}
  {\bf 13} 31--62

\bibitem{corr}
Ankirchner S and Heyne G 2012 {\em Finance and Stochastics\/} {\bf 16} 17--43

\bibitem{emmer}
van Emmerich C 2006 Modelling correlation as a stochastic process Tech. rep.
  Bergische Universit{\"a}t Wuppertal

\bibitem{JunMa2009}
Ma J 2009 {\em Annals of Economics and Finance\/} {\bf 10} 303--327

\bibitem{zomms}
Chetalova D, Sch{\"a}fer R and Guhr T 2015 {\em Journal of Statistical
  Mechanics: Theory and Experiment\/} {\bf 2015} P01029

\end{thebibliography}

\providecommand{\newblock}{}

\end{document}